\documentclass[sigconf,natbib=true,anonymous=False]{acmart}
\usepackage{graphicx}
\usepackage{subfigure}
\usepackage{amsfonts}
\usepackage{amsmath}
\usepackage{booktabs}
\usepackage{multirow}
\usepackage[normalem]{ulem}
\useunder{\uline}{\ul}{}
\usepackage{enumitem}
\usepackage{color}

\usepackage{algpseudocode}
\usepackage[ruled,linesnumbered]{algorithm2e}

\AtBeginDocument{%
  \providecommand\BibTeX{{%
    \normalfont B\kern-0.5em{\scshape i\kern-0.25em b}\kern-0.8em\TeX}}}

\copyrightyear{2023} 
\acmYear{2023} 
\setcopyright{acmlicensed}\acmConference[WWW '23]{Proceedings of the ACM
Web Conference 2023}{May 1--5, 2023}{Austin, TX, USA}
\acmBooktitle{Proceedings of the ACM Web Conference 2023 (WWW '23), May
1--5, 2023, Austin, TX, USA}
\acmPrice{15.00}
\acmDOI{10.1145/3543507.3583513}
\acmISBN{978-1-4503-9416-1/23/04}

\begin{document}
\title{Denoising and Prompt-Tuning for Multi-Behavior Recommendation}

\author{Chi Zhang}
\affiliation{%
  \institution{Harbin Engineering University}
  \city{Harbin}
  \country{China}
}
\email{zhangchi20@hrbeu.edu.cn}

\author{Rui Chen*}
\affiliation{%
  \institution{Harbin Engineering University}
  \city{Harbin}
  \country{China}
}
\email{ruichen@hrbeu.edu.cn}

\author{Xiangyu Zhao*}
\affiliation{%
  \institution{City University of Hong Kong}
  \country{Hong Kong}
}
\email{xianzhao@cityu.edu.hk}

\author{Qilong Han*}
\affiliation{%
  \institution{Harbin Engineering University}
  \city{Harbin}
  \country{China}
}
\email{hanqilong@hrbeu.edu.cn}

\author{Li Li}
\affiliation{%
  \institution{University of Delaware}
  \city{Newark}
  \country{United States}
}
\email{lilee@udel.edu}

\renewcommand{\shortauthors}{Chi Zhang et al.}
\thanks{*Corresponding authors.}

\begin{abstract}
In practical recommendation scenarios, users often interact with items under multi-typed behaviors (e.g., click, add-to-cart, and purchase). Traditional collaborative filtering techniques typically assume that users only have a single type of behavior with items, making it insufficient to utilize complex collaborative signals to learn informative representations and infer actual user preferences. Consequently, some pioneer studies explore modeling multi-behavior heterogeneity to learn better representations and boost the performance of recommendations for a target behavior. However, a large number of auxiliary behaviors (i.e., click and add-to-cart) could introduce irrelevant information to recommenders, which could mislead the target behavior (i.e., purchase) recommendation, rendering two critical challenges: (i) denoising auxiliary behaviors and (ii) bridging the semantic gap between auxiliary and target behaviors. 
Motivated by the above observation, we propose a novel framework--\textbf{\underline{D}}enoising and \textbf{\underline{P}}rompt-\textbf{\underline{T}}uning (DPT) with a three-stage learning paradigm to solve the aforementioned challenges. In particular, DPT is equipped with a pattern-enhanced graph encoder in the first stage to learn complex patterns as prior knowledge in a data-driven manner to guide learning informative representation and pinpointing reliable noise for subsequent stages. Accordingly, we adopt different lightweight tuning approaches with effectiveness and efficiency in the following stages to further attenuate the influence of noise and alleviate the semantic gap among multi-typed behaviors. Extensive experiments on two real-world datasets demonstrate the superiority of DPT over a wide range of state-of-the-art methods. The implementation code is available online
at \url{https://github.com/zc-97/DPT}.
\end{abstract}



\maketitle

\section{Introduction}
\label{sec:introduction}
Recommender systems have been widely used in various online applications (e.g., e-commerce~\cite{ZZS18}, news~\cite{ZZST19}, and social media~\cite{RLL17,ZQD19}) to alleviate information overload and meet user personalized preferences. Traditional collaborative filtering (CF) techniques~\cite{RFZ09,YHC18,WHW19,HDW20,MZX21,WWF21} typically assume that users only interact with items under a single type of behavior, which is insufficient to learn behavior heterogeneity for making accurate recommendations, leading to the problem of \textit{multi-behavior recommendation}.

\begin{figure}[t]
\centering
\setlength{\abovecaptionskip}{1mm}
  \includegraphics[width=\linewidth]{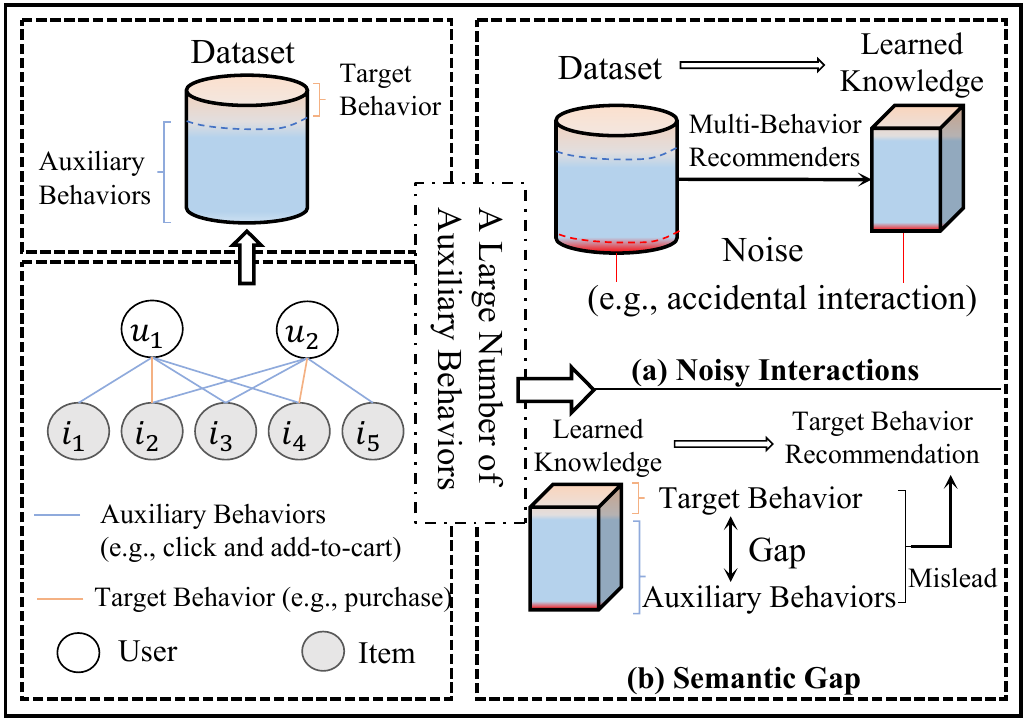}
\caption{An illustration of the two limitations.}
\vspace{-6mm}
\label{fig:motivation}
\end{figure}

In practice, users often interact with items under multi-typed behaviors (e.g., click, add-to-cart, and purchase). Consequently, the mainstream multi-behavior recommendation studies leverage different deep models (e.g., neural collaborative filtering~\cite{GHG19}, attention mechanisms~\cite{GHJ19, XHX20, XHX21, WWQ22} and graph neural networks~\cite{JGH20, CZZ20, XXH21, H21, WHX22}) to learn complex relations from numerous interactions under \textit{auxiliary behaviors} (i.e., click and add-to-cart). Thus, the learned knowledge could serve as additional information to enhance the relatively sparse \textit{target behavior} (i.e., purchase) and better infer users' preferences for target behavior recommendation. Despite existing methods' effectiveness of enhancing target behavior recommendation via multi-behavior heterogeneity, ignoring numerous auxiliary behaviors could also introduce noisy information and irrelevant semantics, leading to inherent sub-optimal representation learning for target behavior recommendation. To this end, as illustrated in Figure~\ref{fig:motivation}, the numerous uncontrollable auxiliary behaviors could render two non-trivial challenges:

\noindent\textbf{Noisy Interactions under Auxiliary Behaviors.} 
Auxiliary behaviors (i.e., click) typically contain some \textit{inherently} noisy user-item interactions (e.g., accidental interactions~\cite{TLF19, WWQ22}), which cannot accurately reflect user interests.
Without fine control, the learned multi-behavior knowledge could be inevitably affected by noise in such data. Therefore, when transferring such knowledge to target behavior recommendation, the uncontrollable noise's influence would be further magnified with increasing auxiliary behaviors in a dataset (i.e., the larger the difference of the sizes of auxiliary and target behaviors, the greater the influence is). Despite its practical value, lacking supervised labels to indicate noisy interactions leads to a unique challenge in solving the denoising problem.

\noindent\textbf{Semantic Gap among Multi-Typed Behaviors.}
While some user-item interactions overlap under multi-typed behaviors, the target behavior still significantly differs from auxiliary behaviors from a semantic feature perspective. For example, a large number of clicks cannot lead to purchases in e-commerce ~\cite{XLG20,WZW20}. Therefore, numerous auxiliary behaviors could inevitably make the learned knowledge over-squash into the semantic space of auxiliary behaviors. Thus, the inherent challenge of bridging the semantic gap between auxiliary and target behaviors lies in how to effectively transfer sufficient target-specific semantics from learned knowledge of multi-typed behaviors to target recommendation without jeopardizing knowledge informativeness.

\noindent\textbf{Contribution.} 
In view of the above challenges, we propose a novel framework--\textbf{\underline{D}}enoising and \textbf{\underline{P}}rompt-\textbf{\underline{T}}uning (DPT) with a \textit{three-stage} learning paradigm for multi-behavior recommendation. Specifically, we derive a pattern-enhanced graph encoder in the first stage to learn complex patterns in a data-driven manner. The patterns are served as prior knowledge to learn informative representations and guide the denoising module to pinpoint inherent noise in auxiliary behaviors. In the second stage, we adopt a re-initializing technique to attenuate the influence of noise further. Finally, we adopt a deeply continuous prompt-tuning paradigm to alleviate the semantic gap among multi-typed behaviors, thus adaptively extracting target-specific information for target behavior recommendation. 
We provide a case study in Section~\ref{subsec:case_study} to show how the three-stage learning paradigm affects recommendations.

We summarize our main technical contributions as follows.
\begin{itemize}[leftmargin=*]
    \item To the best of our knowledge, this is the first paper that proposes to ameliorate multi-behavior recommendation from \textcolor{black}{a new perspective that attenuates auxiliary behaviors' negative influences on target behavior recommendation}. We identify two critical challenges in existing multi-behavior recommendation studies rendered by the numerous auxiliary behaviors, which have been less explored in literature.
    \item We propose a novel multi-behavior recommendation framework named DPT, which is powered by a three-stage learning paradigm to attenuate the influence of noise and alleviate the semantic gap among multi-typed behaviors for target behavior recommendation. In particular, we pinpoint noisy interactions without requiring supervised signals (i.e., labels to indicate noise). We also effectively and efficiently bridge the semantic gap between auxiliary and target behaviors without requiring additional prior knowledge (e.g., knowledge graph). 
    \item We perform extensive experiments on two real-world public recommendation datasets and show that our DPT model consistently outperforms a large number of state-of-the-art competitors.
\end{itemize}

\section{Preliminaries}
\label{sec:preliminaries}
Let $\mathcal{U}=\{u_1,u_2,\cdots,u_{\vert \mathcal{U} \vert}\}$ and $\mathcal{I}=\{i_1,i_2,\cdots,i_{\vert \mathcal{I} \vert}\}$ denote the set of users and items, respectively. Following previous studies~\cite{JGH20, CZZ20, XXH21, WHX22}, we take purchase as the target behavior and others serving as auxiliary behaviors (i.e., click, add-to-favorite, and add-to-cart). For simplicity, \textcolor{black}{we use $b^a$ and $b^t$ to mean auxiliary behaviors and the target behavior in the remainder, respectively}. Accordingly, we define the multi-behavior data as a set of interaction matrices $\mathcal{A} = \{A^a, A^t\}$, where $A^a, A^t \in\mathbb{R}^{\vert \mathcal{U} \vert \times \vert \mathcal{I} \vert}$. \textcolor{black}{Each element in $A^*$ indicates whether a user $u$ has interacted with item $i$ under behavior $*$ (i.e., $A^*_{ui} = 1$) or not (i.e., $A^*_{ui} = 0$), where $* \in \{a,t\}$}.

\noindent\textbf{User-Item Multi-Behavior Graph.} 
\textcolor{black}{We construct $\mathcal{G}=\{\mathcal{G}^a, \mathcal{G}^t\}$ to represent interactions under auxiliary and target behaviors.} For each graph $\mathcal{G}^*=(\mathcal{V}, \mathcal{E}^*) \in \mathcal{G}$, there is an edge $\varepsilon^*_{ui}$ between user $u$ and item $i$ in $\mathcal{E}^*$ iff $A^*_{ui} = 1$, where $\mathcal{V} = \mathcal{U} \cup \mathcal{I}$ and $* \in \{a,t\}$. 

\noindent\textbf{Problem Statement.} We formally describe our task as follows: 
\textbf{Input}: constructed multi-behavior graph $\mathcal{G}$. \textbf{Output}: a noiseless multi-behavior graph $\mathcal{G}'=\{\mathcal{{G}}'^a, \mathcal{G}^t\}$, where $\mathcal{{G}}'^a = \{ \mathcal{V}, \mathcal{E}'^a \}$ and $\mathcal{E}'^a \subseteq \mathcal{E}^a$, and the predictive function $\mathcal{F}(u,i\vert \mathcal{G}', \Theta)$, which estimates the likelihood of user $u$ adopting the item $i$ of the target behavior type, where $\Theta$ is the set of model parameters.

\section{Methodology}
\label{sec:methodology}

\begin{figure*}[t]
  \centering
  \setlength{\abovecaptionskip}{1mm}
    \includegraphics[width=\linewidth]{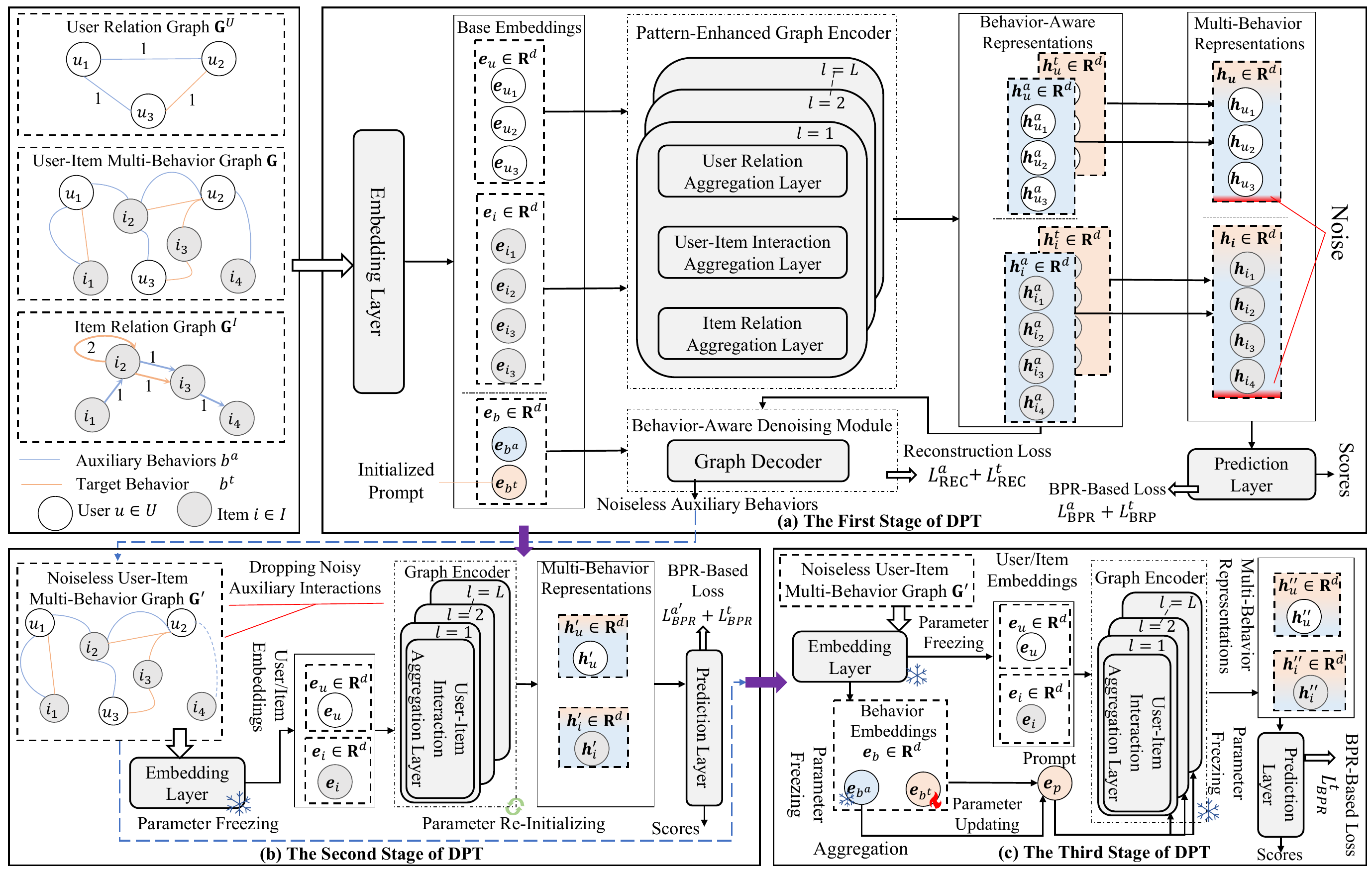}
  \caption{The architecture of the proposed DPT model. \textcolor{black}{ (a) The first stage of DPT takes different relation graphs as input and outputs a noiseless graph $\mathcal{G}'$ for subsequent stages; (b) The second stage of DPT re-initializes partially learned parameters to further attenuate noise's influence; (c) The third stage of DPT uses continuous prompts to bridge the behavioral semantic gap.}
  }
 \label{fig:model}
\vspace{-3mm}
\end{figure*}

As illustrated in Figure~\ref{fig:model}, DPT is learned in a three-stage learning paradigm. Specifically, the first stage is equipped with a pattern-enhanced graph encoder to learn informative representations and guide the denoising module to generate a noiseless multi-behavior graph for subsequent stages. After that, we adopt different lightweight tuning approaches to further attenuate noise's influence and alleviate the semantic gap among multi-typed behaviors.

\subsection{User/Item Relation Graph Construction}
\label{subsec:graphs}
Recall that lack of labels for noise identification makes the denoising problem challenging. 
In practice, there typically exists multiple patterns between user-user relations (e.g., social relations~\cite{WLT21}) and item-item relations (e.g., knowledge-driven relations~\cite{WZZ19, WZM20}), which could serve as prior knowledge to improve the representation learning ability. Hence, we construct two types (i.e., user and item) of relation graphs to mine patterns in a data-driven manner.

\subsubsection{User Relation Graph.} In practice, user-side information could be well reflected by their co-interactions~\cite{HXX21, TLZ22}. Consequently, we utilize users' co-interactive behaviors (e.g., co-view and co-purchase) to learn the complex patterns. Mathematically, there is an edge $\varepsilon^{U,*}_{uv}$ between user $u,v\in\mathcal{U}$ in the user relation graph $\mathcal{G}^U$ under the behavior $*$ iff $\mathcal{E}^*_{u\cdot} \cap \mathcal{E}^*_{v\cdot} \neq \emptyset \:\textrm{and}\: \mathcal{E}^*_{u\cdot} \cup \mathcal{E}^*_{v\cdot} \neq \emptyset.$
The weight $w^{U,*}_{uv}$ of $\varepsilon^{U,*}_{uv}$ is calculated by Jaccard similarity as $\frac{\vert \mathcal{E}^*_{u\cdot} \cap \mathcal{E}^*_{v\cdot} \vert}{\vert \mathcal{E}^*_{u\cdot} \cup \mathcal{E}^*_{v\cdot} \vert}$ to indicate the relevance strength between the users.

\subsubsection{Item Relation Graph.} As suggested by previous studies~\cite{XHX21, HZC22}, unlike user relations, items possess directed relations due to the existence of temporal information. Consequently, we construct a directed item relation graph for better representation learning. Specifically, we sort each user $u$'s interactions into a sequence $S_u$ according to interaction timestamps, and count the number of occurrences of each pair of items $(i,j)$ in a particular order (e.g., $i\xrightarrow{*} j$) from all sequences. Formally, there is an edge $\varepsilon^{I,*}_{ij}$ from item $i$ to $j$ in the item relation graph $\mathcal{G}^I$ under behavior $*$ iff $\text{Cnt}(i\xrightarrow{*} j)>0$, where $\text{Cnt}(\cdot)$ is a counter. The weight $w^{I,*}_{ij}$ of $\varepsilon^{I,*}_{ij}$ is calculated by Jaccard similarity as $\frac{\text{Cnt}(i\xrightarrow{*} j)}{\text{Cnt}(i\xrightarrow{*} j)+\text{Cnt}(j\xrightarrow{*} i)}$ to indicate the sequential strength.


\subsection{Embedding Layer}
\label{subsec:embedding layer}
To train DPT, we first learn the embeddings of users, items, and behaviors by mapping their IDs to dense embedding vector $\boldsymbol{e} \in \mathbb{R}^d$. Formally, we build four embedding look-up tables for initialization:
\begin{equation}
    \boldsymbol{e}_{u} = \boldsymbol{x}_{u}\boldsymbol{W}_{u} \: ; \:
    \boldsymbol{e}_{i} = \boldsymbol{x}_{i}\boldsymbol{W}_{i} \: ; \:
    \boldsymbol{e}_{b^a} = \boldsymbol{x}_{b^a}\boldsymbol{W}_{b^a} \: ; \:
    \boldsymbol{e}_{b^t} = \boldsymbol{x}_{b^t}\boldsymbol{W}_{b^t},
    \label{eq:embedding}
\end{equation}
where $\boldsymbol{W}_{u} \in \mathbb{R}^{\vert \mathcal{U} \vert \times d}$, $\boldsymbol{W}_{i} \in \mathbb{R}^{\vert \mathcal{I} \vert \times d}$, $\boldsymbol{W}_{b^a} \in \mathbb{R}^{(K-1)\times d} $, and $\boldsymbol{W}_{b^t} \in \mathbb{R}^{d}$ are trainable matrices. $\boldsymbol{x}_{u}$, $\boldsymbol{x}_{i}$, $\boldsymbol{x}_{b^a}$ and $\boldsymbol{x}_{b^t}$ are the one-hot encoding vectors of the IDs of user $u$, item $i$, auxiliary behavior $b^a$, and target behavior $b^t$, respectively. 


\subsection{Pattern-Enhanced Graph Encoder}
\label{subsec:graph encoderr}
\begin{figure}[t]
  \centering
  \setlength{\abovecaptionskip}{1mm}
  \includegraphics[width=\linewidth]{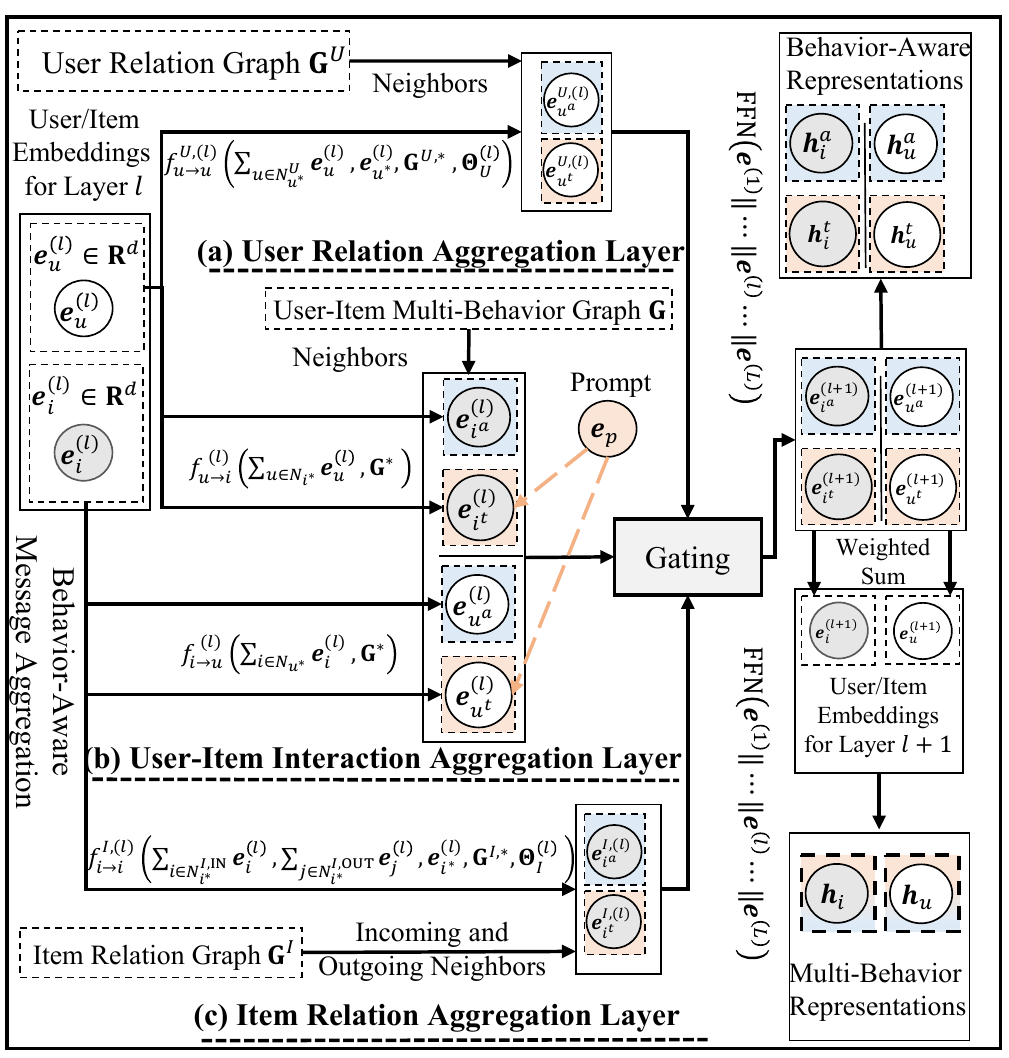}
  \caption{The proposed pattern-enhanced graph encoder.}
 \label{fig:encoder}
\end{figure}

As illustrated in Figure~\ref{fig:encoder}, we design a pattern-enhanced graph encoder, consisting of three disentangled aggregation layers to individually learn multi-view patterns based on constructed graphs.

\subsubsection{User Relation Aggregation Layer.} Since the user relation graph $\mathcal{G}^U$ is an undirected graph, we encode such relations without distinguishing directions. Formally, for each user $u^*$, we aggregate the information from his/her neighbor's information and himself/herself as per $\mathcal{G}^U$ to learn similar behavior patterns among users and generate layer-specific behavior-aware encoded vectors as follows
\begin{equation}
    \boldsymbol{e}^{U,(l)}_{u^*} = f^{U,(l)}_{u\rightarrow u}(\sum_{u\in N^U_{u^*}}\boldsymbol{e}^{(l)}_u, \boldsymbol{e}^{(l)}_{u^*}, \mathcal{G}^{U,*}, \Theta^{(l)}_U),
\label{eq:user relation aggregation layer}
\end{equation}
where $f^{U,(l)}_{u \rightarrow u}(\cdot)$ is the aggregation function to encode user relations, $l\in \mathbb{R}$ is the number of layer, $N^U_{u^*}$ is the neighbor of $u^*$ in $\mathcal{G}^{U,*}$ under behavior $*\in\{a,t\}$, and $\Theta^{(l)}_U$ is the set of trainable parameters. Motivated by~\cite{DTD17, HZC22}, which utilize a convolution operator to preserve dimension-wise information and effectively encode homogeneous relations, we implement $f^{U,(l)}(\cdot)$ and generate $\boldsymbol{e}^{U,(l)}_{u^*}$ via
\begin{equation}
    \boldsymbol{e}^{U,(l)}_{u^*} = \text{Conv}^{U,(l)}( [\sum_{u\in N^U_{u^*}}{(w^{U,*}_{uu^*} \boldsymbol{e}^{(l)}_u)} \Vert \boldsymbol{e}^{(l)}_{u^*}], \Bar{\Theta}^{(l)}_U),
\label{eq:user aggregation}
\end{equation}
where $\text{Conv}^{U,(l)}(\cdot)$ is a layer-specific convolution operator with stride 1 and filter size $2 \times 1$, $w^{U,*}_{uu^*}$ is the normalized weight of edge $\varepsilon^{U,*}_{uu^*}$ in $\mathcal{G}^{U,*}$, $\Vert$ is the concatenation operation, and $\Bar{\Theta}^{(l)}_U$ is the set of trainable weight parameters of the convolution operator.

\subsubsection{Item Relation Aggregation Layer.} Since items typically have directed patterns, we perform an attention-based message passing scheme~\cite{VCC18} (i.e., $f^{I,(l)}_{i \rightarrow i}(\cdot)$) for adaptively encoding. For each item $i^*$ in the item relation graph $\mathcal{G}^{I,*}$ under behavior $*$, we first assign a $2$-dimensional attention vector $\boldsymbol{\alpha}^{*}$ to weight $i^*$'s incoming (i.e., $i \in N^{I,\text{IN}}_{i^*}$) and outgoing (i.e., $j \in N^{I,\text{OUT}}_{i^*}$) neighbors. Formally, we implement $f^{I,(l)}_{i \rightarrow i}(\cdot)$ and generate $\boldsymbol{e}^{I,(l)}_{i^*}$ as follows
\begin{equation}
    \begin{aligned}
    &\boldsymbol{e}^{I,(l)}_{i^*} = f^{I,(l)}_{i \rightarrow i}(\sum_{i\in N^{I,\text{IN}}_{i^*}}\boldsymbol{e}^{(l)}_i, \sum_{j\in N^{I,\text{OUT}}_{i^*}}\boldsymbol{e}^{(l)}_j, \boldsymbol{e}^{(l)}_{i^*}, \mathcal{G}^{I,*}, \Theta^{(l)}_I),\\
    &\boldsymbol{\alpha}^{*} = [\alpha^{*}_{\text{IN}}, \alpha^{*}_{\text{OUT}}] =  \frac{\text{exp}(\boldsymbol{e}^{\mathsf{T}}_{i^*}\boldsymbol{W}^{(l)}_{I,n}/\sqrt{d})}
    {\sum_{n\in\{\text{IN},\text{OUT}\}}\text{exp}(\boldsymbol{e}^{\mathsf{T}}_{i^*}\boldsymbol{W}^{(l)}_{I,n}/\sqrt{d}))  }  ,\\
    &\boldsymbol{e}^{(l)}_{N^I_{i^*}} = \alpha^{*}_{\text{IN}}\sum_{{i \in N^{I,\text{IN}}_{i^*}}}{(w^{I,*}_{ii^*} \boldsymbol{e}^{(l)}_i)} + \alpha^{*}_{\text{OUT}} \!\!\!\!\!\! \sum_{i^* \in N^{I,\text{OUT}}_{i^*}}{(w^{I,*}_{i^*j} \boldsymbol{e}^{(l)}_j)},\\
    &\boldsymbol{e}^{I,(l)}_{i^*} = \text{Conv}^{I,(l)}(\boldsymbol{e}^{(l)}_{N^I_{i^*}} \Vert \boldsymbol{e}_{i^*}^{(l)}, \Bar{\Theta}^{(l)}_{I}),
    \end{aligned}
\label{eq: item relation aggregation layer}
\end{equation}
where $\text{Conv}^{U,(l)}(\cdot)$ is a similar convolution used in Eq.~(\ref{eq:user aggregation}) with different parameters, $w^{I,*}_{ii^*}, w^{I,*}_{i^*j}$ are the normalized weights of edge in $\mathcal{G}^{I,*}$, and $\Theta^{(l)}_{I}$ is the set of trainable parameters, including attention matrices $W^{(l)}_{I,n}\in\mathbb{R}^{d\times d}$ and the convolutional filter $\Bar{\Theta}^{(l)}_{I}$ .

\subsubsection{User-Item Interaction Aggregation Layer.} 
For each user $u^*$ and item $i^*$, we formally present a lightweight message passing scheme based on LightGCN~\cite{HDW20} as follows
\begin{equation}
\begin{aligned}
    \boldsymbol{e}^{(l)}_{u^*} = f^{(l)}_{i\rightarrow u}(\sum_{i\in N_{u^*}}\boldsymbol{e}^{(l)}_i, \mathcal{G}^{*}) \: &; \: \boldsymbol{e}^{(l)}_{i^*} = f^{(l)}_{u\rightarrow i}(\sum_{u\in N_{i^*}}\boldsymbol{e}^{(l)}_u, \mathcal{G}^{*}),\\
    \boldsymbol{e}^{(l)}_{u^*} = \sum_{i\in N_{u^*}}\boldsymbol{e}^{(l)}_i \: &; \: \boldsymbol{e}^{(l)}_{i^*} = \sum_{u\in N_{i^*}}\boldsymbol{e}^{(l)}_u.
\end{aligned}
\label{eq:user-item aggregation}
\end{equation}

To encoder multiple patterns from different graphs into embeddings, we design a gating operator, which could adaptively balance and fuse the different views generated from the $(l)$- to $(l+1)$-th layers. More specifically, we generate the pattern-enhanced user/item embedddings $\boldsymbol{e}^{(l+1)}_{u^*}$ and $\boldsymbol{e}^{(l+1)}_{i^*}$ under the type of behavior $*$ via
\begin{equation}
\begin{aligned}
    \beta^{*}_{u} = \sigma([\boldsymbol{e}^{(l)}_{u^*}\Vert \boldsymbol{e}^{U,(l)}_{u^*} ]\boldsymbol{W}^{(l)}_{UU}) \: &; \:
    \beta^{*}_{i} = \sigma([\boldsymbol{e}^{(l)}_{i^*}\Vert \boldsymbol{e}^{I,(l)}_{i^*} ]\boldsymbol{W}^{(l)}_{II}), \\
    \boldsymbol{e}^{(l+1)}_{u^*} =  \boldsymbol{e}^{(l)}_{u^*} + \frac{1-\beta^{*}_{u}}{\beta^{*}_{u}} \boldsymbol{e}^{U,(l)}_{u^*} &;
    \boldsymbol{e}^{(l+1)}_{i^*} = \boldsymbol{e}^{(l)}_{i^*} + \frac{1-\beta^{*}_{i}}{\beta^{*}_{i}} \boldsymbol{e}^{I,(l)}_{i^*},
\end{aligned}
\label{eq:behavior-aware embeddings}
\end{equation}
where $\sigma(\cdot)$ is the sigmoid function, $\beta^{*}_{u}, \beta^{*}_{i}\in\mathbb{R}$ are weight scalars, and $\boldsymbol{W}^{(l)}_{UU},\boldsymbol{W}^{(l)}_{II}\in\mathbb{R}^{2d\times1}$ are trainable parameters. 
Accordingly, we generate multi-behavior embeddings in the $(l+1)$-th layer via
\begin{equation}
    \boldsymbol{e}^{(l+1)}_{u} = \frac{1}{\vert\{a,t\}\vert}\sum_{* \in \{a,t\}}{\boldsymbol{e}^{(l)}_{u^*}}\: ; \: 
    \boldsymbol{e}^{(l+1)}_{i} = \frac{1}{\vert\{a,t\}\vert}\sum_{* \in \{a,t\}}{\boldsymbol{e}^{(l)}_{i^*}}.
\label{eq:multi-behavior embeddings}
\end{equation}
We concatenate user/item embeddings across each layer to generate the final behavior-aware (i.e., $\boldsymbol{h}^*_{u},\boldsymbol{h}^*_{i}\in\mathbb{R}^d$) and multi-behavior (i.e., $\boldsymbol{h}_{u},\boldsymbol{h}_{i}\in\mathbb{R}^d$) representations:
\begin{equation}
\begin{aligned}
    \boldsymbol{h}^*_{u} = f_U(\boldsymbol{e}^{(1)}_{u^*}\Vert\boldsymbol{e}^{(2)}_{u^*}\Vert\cdots\Vert\boldsymbol{e}^{(L)}_{u^*});    \boldsymbol{h}^*_{i} = f_I(\boldsymbol{e}^{(1)}_{i^*}\Vert\boldsymbol{e}^{(2)}_{i^*}\Vert\cdots\Vert\boldsymbol{e}^{(L)}_{i^*}),\\
    \boldsymbol{h}_{u} = f_U(\boldsymbol{e}^{(1)}_{u}\Vert\boldsymbol{e}^{(2)}_{u}\Vert\cdots\Vert\boldsymbol{e}^{(L)}_{u});    \boldsymbol{h}_{i} = f_I(\boldsymbol{e}^{(1)}_{i}\Vert\boldsymbol{e}^{(2)}_{i}\Vert\cdots\Vert\boldsymbol{e}^{(L)}_{i}),
\end{aligned}
\label{eq:final representations}
\end{equation}
where $f_U$ and $f_I$ are feedforward layers activated by the ReLU function with different trainable parameters $\boldsymbol{W}_U,\boldsymbol{W}_I\in\mathbb{R}^{(L\times d)\times d}$. 

\subsection{Behavior-Aware Denoising Module}
\label{subsec:denoising}
In multi-behavior graph $\mathcal{G}$, normal interactions usually exhibit high consistency with the graph structure from a local-global unified perspective, while noisy interactions do not. Thus, once we learn the pattern-enhanced user/item representations to parameterize $\mathcal{G}$, it should be difficult to reconstruct noisy interactions from such informative representations. It follows that we can compare difficulty levels of information reconstruction to pinpoint inherent noise.

\subsubsection{Graph Decoder.} We design a behavior-aware graph decoder as a discriminator to perform information reconstruction. It takes the learned behavior-aware representations as input and generates the probability graph via
\begin{equation}
    \boldsymbol{h}_{\mathcal{G}^*} = \sigma(\boldsymbol{H}^*_U \boldsymbol{e}^{\mathsf{T}}_{b^*} \boldsymbol{e}_{b^*} {\boldsymbol{H}^*_I}^{\mathsf{T}}),
\label{eq:decoder}
\end{equation}
where $\boldsymbol{h}_{\mathcal{G}^*}\in [0,1]^{\vert\mathcal{U}\vert\times\vert\mathcal{I}\vert}$ is the parameterized graph under behavior $*$, $\boldsymbol{H}^*_U\in\mathbb{R}^{\vert\mathcal{U}\vert\times d}, \boldsymbol{H}^*_I\in\mathbb{R}^{\vert\mathcal{I}\vert\times d}$ are the learned representations of the user/item set, and $\boldsymbol{e}_{b^*}\in\mathbb{R}^d$ is the behavior embedding. Thus, we formulate the information reconstruction task as a binary classification task to predict user-item interactions in graph $\mathcal{G}$:
\begin{equation}
\begin{aligned}
    &\mathcal{L}^*_{\text{REC}} = -\frac{1}{\vert \mathcal{U}\vert\vert \mathcal{I} \vert}\sum_{u\in\mathcal{U},i\in\mathcal{I}}(\mathcal{G}^*_{ui}\text{log}(\boldsymbol{h}_{\mathcal{G}^*_{ui}})+(1-\mathcal{G}^*_{ui})\text{log}(1-\boldsymbol{h}_{\mathcal{G}^*_{ui}})),\\
    &\mathcal{L}_{\text{REC}} = \frac{1}{\vert\{a,t\}\vert}\sum_{*\in\{a,t\}}\mathcal{L}^*_{\text{REC}}.
\end{aligned}
\label{eq:L_rec}
\end{equation}
Note that we also enhance the decoder (i.e., Eq.~(\ref{eq:decoder})) to reconstruct the target behavior graph since target behavior (e.g., purchase) is more reliable than auxiliary behaviors (e.g., click) in real-world scenarios and can help avoid incorrect noise identification. Moreover, it is reasonable to assume that most interactions should be noiseless. Therefore, we could gradually learn informative interactions' distribution by minimizing Eq.~(\ref{eq:L_rec}). In this case, interactions with higher loss scores can be identified as noisy interactions.

\subsection{Three-Stage Learning Paradigm}
Based on the above components, we can obtain the learned pattern-enhanced representations and noiseless auxiliary behaviors. However, the learned parameters in the encoder may be affected by inherent noise in the original data, and the semantic gap may still exist, leading to sub-optimal knowledge learning. Consequently, we propose a \textit{three-stage learning paradigm} in DPT. In the first stage, DPT leverages pattern-enhanced representations to guide noise identification for the following stages. In the second stage, DPT adopts a re-initializing method to attenuate the noise's influence on learned knowledge. In the third stage, DPT integrates target-specific prompt-tuning to bridge the semantic gap.

\subsubsection{The First Stage of DPT}
\label{subsubsec:first stage}
As shown in Figure~\ref{fig:model}(a), DPT takes the constructed relation graphs as input and encodes patterns into embeddings, which helps pinpoint noise. Specifically, we generate base embeddings by Eq.~(\ref{eq:embedding}) and input them to the pattern-enhanced graph encoder (i.e., Section~\ref{subsec:graph encoderr}) to encode multi-view patterns so as to generate behavior-aware and multi-behavior representations by Eq.~(\ref{eq:final representations}).
We leverage the obtained behavior-aware representations to guide the denoising module to learn parameterized graph $\boldsymbol{h}_{\mathcal{G}^*}$ by Eq.~(\ref{eq:decoder}) and calculate reconstruction loss $\mathcal{L}_{\text{REC}}$ by Eq.~(\ref{eq:L_rec}). We convert $\boldsymbol{h}_{\mathcal{G}^a}$ into a hard-coding binary (i.e., 0 vs. 1) distribution, which indicates whether the interaction under auxiliary behaviors is noisy or not, to generate the denoised multi-behavior graph $\mathcal{G}'$. Here we could adopt differentiable methods (e.g., Gumbel-softmax function~\cite{WYW19, YGH19, ZLL21, QWL21}) to binarize the real-valued graph $\boldsymbol{h}_{\mathcal{G}^a}$ in each training iteration. However, reconstructing and generating such a large-scale graph according to Eq.~(\ref{eq:decoder}) and Eq.~(\ref{eq:L_rec}) leads to large computational costs. 
Motivated by sub-graph sampling~\cite{XHX21,ZMG21} and hyper-parameter optimizing~\cite{LSY19,YWW21} strategies, we first sample a mini-batch $\mathcal{B}$ of user/item interactions to optimize Eq.~(\ref{eq:L_rec}), thus endowing the denoising module with the ability of handling large-scale graphs. Then we leverage the optimized parameters to generate the probability graph instead of generating it iteratively. Specifically, we adopt a simple threshold strategy (i.e., $\boldsymbol{h}_{\mathcal{G}^a_{ui}}=0$ if $\boldsymbol{h}_{\mathcal{G}^a_{ui}}<0.5-\delta$, $\boldsymbol{h}_{\mathcal{G}^a_{ui}}=1$ otherwise) to binarize $\boldsymbol{h}_{\mathcal{G}^a}$, where $\delta$ is a disturber to control reliability of the denoising discriminator. Note that $\delta\rightarrow0^+$ and $\delta\rightarrow0.5^-$ lead to more and less identified noise, respectively. After that, we can generate $\mathcal{G}' = \{\mathcal{G}^a \odot \boldsymbol{h}_{\mathcal{G}^a},\mathcal{G}^{t}\}$ for the following stages of DPT, where $\odot$ is the element-wise dot product operator. 

However, noise identification highly depends on the reliability of the graph encoder. A pure unsupervised task for denoising may make the optimization process irrelevant to multi-behavior recommendation. Inspired by the co-guided learning scheme~\cite{ZXY22}, we incorporate a multi-behavior recommendation task into the first stage, and thus can avoid unreliable learning (e.g., false noise identification) in the early stage of the training process. Specifically, we formulate the multi-behavior recommendation task as minimizing the Bayesian Personalized Ranking (BPR) objective function:
\begin{equation}
\begin{aligned}
    &\mathcal{L}_{BPR} = \frac{1}{\vert\{a,t\}\vert\vert \mathcal{B} \vert}\sum_{*\in\{a,t\}}\sum_{(u,i^*,j*)\in\mathcal{B}}\mathcal{L}^*_{\text{BRP}}(u,i^*,j^*),\\
    &\mathcal{L}^*_{\text{BRP}}(u,i^*,j^*) = -\log(\sigma(\text{sim}(\boldsymbol{h}_u, \boldsymbol{h}_{i^*}) - \text{sim}(\boldsymbol{h}_u, \boldsymbol{h}_{j^*}))),
\end{aligned}
\label{eq:first stage BPR}
\end{equation}
where $\text{sim}(\cdot)$ is a similarity function (e.g., inner product or a neural network), $u\in\mathcal{U}$, $i^*\in\mathcal{I}$, $\mathcal{G}^*_{ui^{*}}=1$, and $j^*$ is a randomly sampled item from $\mathcal{I}$ with $\mathcal{G}^*_{uj^{*}}=0$ in each mini-batch $\mathcal{B}$.


\subsubsection{The Second Stage of DPT} While representations learned in the first stage are informative, noise still inevitably affects the learned model parameters (e.g., the embedding layer) because the representation-based denoising module can capture only reliable, but not all, noise. Intuitively, we could leverage $\mathcal{G}'$ to retrain (e.g., full-tuning) the entire model. However, it is less desirable due to the large additional computational costs of model re-training. In addition, the optimized parameters in the embedding layer are informative enough to reflect pattern information, and thus there is no need to encode relation graphs in the second stage. Accordingly, we can fine-tune a few parameters instead of full-tuning the entire model to further attenuate noise's influence. Due to the disentangled design choice of aggregation layers, we can efficiently aggregate user-item interactions (i.e., by Eq.~(\ref{eq:user-item aggregation}) and Eq.~(\ref{eq:multi-behavior embeddings})) without repeatedly encoding patterns.
More specifically, as shown in Figure~\ref{fig:model}(b), we freeze the parameters of the embedding layer to generate the base embeddings for users and items. Then, we input them to the user-item interaction aggregation layer (i.e., Eq.~(\ref{eq:user-item aggregation}) and Eq.~(\ref{eq:multi-behavior embeddings})) to iteratively generate multi-behavior embeddings, and learn the final multi-behavior representations (i.e., $\boldsymbol{h}'_u$ and $\boldsymbol{h}'_i$) by Eq.~(\ref{eq:final representations}) with re-initialized parameters (i.e., $\boldsymbol{W}_U,\boldsymbol{W}_I\in\mathbb{R}^{Ld\times d}$). According to the learned representations, we rewrite the BPR loss based on the noiseless graph as follows
\begin{equation}
        \mathcal{L}^*_{\text{BRP}}(u,i^*,j^*) = -\log(\sigma(\text{sim}(\boldsymbol{h}'_u, \boldsymbol{h}'_{i^*}) - \text{sim}(\boldsymbol{h}^{'}_u, \boldsymbol{h}'_{j^*}))),
\label{eq:bpr the second stage}
\end{equation}
where $u\in\mathcal{U}$, $i^*,j^*\in\mathcal{I}$, $\mathcal{G}'_{ui^{*}}=1$, and $\mathcal{G}'_{uj^{*}}=0$. By minimizing Eq.~(\ref{eq:bpr the second stage}), we denote the parameter set as $\Theta_2 = \{\boldsymbol{W}_U,\boldsymbol{W}_I\}$, which contains $2Ld \times d$ parameters. It can be seen that we only tune a small amount of parameters $2Ld \times d \ll (\vert\mathcal{U}\vert + \vert\mathcal{I}\vert) \times d$, instead of tuning the embedding layer.

\subsubsection{The Third Stage of DPT} After the above stages, we can generate noiseless auxiliary behaviors and informative knowledge. However, bridging the semantic gap among multi-typed behaviors is still challenging because it requires transferring sufficient target-specific semantics without jeopardizing learned multi-behavioral knowledge. Inspired by the effectiveness of the prompt-tuning paradigm~\cite{JTC22,XPK22,GLF22,WXY22,SZH22}, we adopt a deep continuous prompt-tuning approach in the third stage to alleviate the semantic gap between auxiliary and target behaviors. Specifically, as illustrated in Figure~\ref{fig:model}(c), we utilize the aggregation of multi-typed behavior embeddings (i.e., $\boldsymbol{e}_{b^a}$ and $\boldsymbol{e}_{b^t}$) to generate a prompt embedding $\boldsymbol{e}_p\in\mathbb{R}^d$, instead of initializing it randomly, to better understand prompt semantics~\cite{SZH22}. Note that, during the initializing process, we freeze auxiliary behaviors' embedding parameters (i.e., $\boldsymbol{W}_{b^a}$) and update the target behavior's parameters (i.e., $\boldsymbol{W}_{b^t}\in\mathbb{R}^d$). Therefore, to adopt prompt-tuning in the third stage, we freeze not only the parameters of the embedding layer (except $\boldsymbol{W}_{b^t}$) but also the parameter set $\Theta_2$ learned in the second stage. More specifically, to generate embeddings of the graph encoder's $(l)$-th layer, we only adopt prompt under the target behavior via
\begin{equation}
\begin{aligned}
        \boldsymbol{e}^{(l)}_{u^t} = f^{(l)}_{i\rightarrow u}(\! \sum_{i\in N_{u^t}} \!\! \boldsymbol{e}^{(l)}_i, \mathcal{G}^{t}, \textcolor{orange}{\boldsymbol{e}_p})&; \boldsymbol{e}^{(l)}_{i^t} = f^{(l)}_{u\rightarrow i}(\! \sum_{u\in N_{i^t}} \!\! \boldsymbol{e}^{(l)}_u, \mathcal{G}^{t}, \textcolor{orange}{\boldsymbol{e}_p}),\\
        \boldsymbol{e}^{(l)}_{u^a} = f^{(l)}_{i\rightarrow u}(\sum_{i\in N_{u^a}}\boldsymbol{e}^{(l)}_i, \mathcal{G}^{a})&; \boldsymbol{e}^{(l)}_{i^t} = f^{(l)}_{u\rightarrow i}(\sum_{u\in N_{i^a}}\boldsymbol{e}^{(l)}_u, \mathcal{G}^{a}),
\end{aligned}
\label{eq:differnt prompt approaces}
\end{equation}
where we can leverage various ways to adopt the prompt, e.g., add, concatenate, and projection operators. Inspired by VPT~\cite{JTC22}, we use the simple yet effective \textit{add} operator to adopt the prompt in each layer. We further conduct experiments in Section~\ref{sec:ablation study} 
over these variants to justify our design choice. We formulate the above process via
\begin{equation}
    \boldsymbol{e}^{(l)}_{u^t} = \textcolor{orange}{\boldsymbol{e}_p} + \sum_{i\in N_{u^t}}\boldsymbol{e}^{(l)}_i  \: ; \: \boldsymbol{e}^{(l)}_{i^t} = \textcolor{orange}{\boldsymbol{e}_p} + \sum_{u\in N_{i^t}}\boldsymbol{e}^{(l)}_u.
\label{eq:add prompt}
\end{equation}
We can leverage the identical process to learn the final multi-behavior representations by Eq.~(\ref{eq:final representations}) and rewrite the BPR loss based on the noiseless graph for the target behavior via
\begin{equation}
    \mathcal{L}_{BPR} = \frac{1}{\vert \mathcal{B} \vert}\sum_{(u,i^t,j^t)\in\mathcal{B}} \!\!\!\!\! - \log(\sigma(\text{sim}(\boldsymbol{h}''_u, \boldsymbol{h}''_{i^t}) - \text{sim}(\boldsymbol{h}''_u, \boldsymbol{h}''_{j^t}))).
\label{eq:third stage BPR}
\end{equation}
We only optimize the objective function under the target behavior, and the prompt only has $d\ll2L\times d \times d$ trainable parameters.

\subsubsection{Model Complexity Analysis} 
We analyze the size of the trainable parameters in DPT at each stage. In this first stage, we update all parameters of the embedding layer and the pattern-enhanced graph encoder, which are denoted by $\Theta_1$. We have $\vert\Theta_1\vert = (\vert \mathcal{U} \vert + \vert \mathcal{I} \vert ) \times d + 4L \times (d \times d + 1)$. In the second stage, we only train a small amount of parameters, which are denoted by $\Theta_2$. We have $\vert \Theta_2 \vert = 2L\times (d \times d)$. In the third stage, we train the parameters of the prompt, denoted by $\Theta_3$. Its size is $\vert \Theta_3 \vert = d$. In conclusion, the proposed DPT can achieve comparable space complexity with state-of-the-art multi-behavior recommendation methods (e.g., CML~\cite{WHX22}). The adopted lightweight approaches (i.e., the second and third stages of DPT) only tune/add a small number of parameters compared with the ones in the embedding layer for encoding users and items (i.e., $\mathcal{U}$ and $\mathcal{I}$ are typically of large sizes).

\section{Evaluation}
\label{sec:experiments}
In this section, we aim to answer the following research questions:
\begin{itemize}[leftmargin=*]
    \item \textbf{RQ1}: Dose the proposed DPT model outperforms other state-of-the-art multi-behavior recommendation methods?
    \item \textbf{RQ2}: How do different stages and behaviors of DPT contribute to the performance of target behavior recommendation?
    \item \textbf{RQ3}: How is the interpretation ability of the three-stage learning paradigm in DPT for denoising and target behavior recommendation?
\end{itemize}

\subsection{Experimental Settings}
\subsubsection{Datasets and Evaluation
Metrics.} To evaluate the effectiveness of the proposed DPT model, we conduct experiments on two public recommendation datasets: (1) \textbf{Tmall} that is
collected from the Tmall E-commerce platform, and (2) \textbf{IJCAI-Contest} that is adopted in IJCAI15 Challenge from a business-to-customer retail system (referred to as IJCAI for short). These datasets have same types of behaviors, including \textit{click},\textit{add-to-favorite},\textit{add-to-cart}, and \textit{purchase}. Identical to previous studies~\cite{JGH20,XHX20,WHX22}, we set the purchase behavior as the target behavior, and others are considered as auxiliary behaviors. Then we filter out users whose interactions are less than 3 under the purchase behavior. Moreover, we adopt the widely used \textit{leave-one-out} strategy by leaving users’ last interacted items under the purchase behavior as the test set. Two evaluation metrics, HR (Hit Ratio) and NDCG (Normalized Discounted Cumulative Gain, N for short) @ 10, are used for performance evaluation. The statistics of the two datasets are summarized in Table~\ref{tab:dataset}.

\begin{table}[]
\caption{Statistics of experimented datasets.}
\label{tab:dataset}
\resizebox{\linewidth}{!}{
\begin{tabular}{@{}ccccc@{}}
\toprule
\textbf{Dataset} &
  \multicolumn{1}{c}{\textbf{User\#}} &
  \multicolumn{1}{c}{\textbf{Item\#}} &
  \multicolumn{1}{c}{\textbf{Interaction\#}} & 
  \multicolumn{1}{c}{\textbf{Interactive Behavior Type}} \\ \midrule
IJCAI  & 17,435 & 35,920 & 799,368  & \{Click,Favorite,Cart,Purchase\} \\
Tmall    & 31,882 & 31,232 & 1,451,219     & \{Click,Favorite,Cart,Purchase\} \\
\bottomrule
\end{tabular}
}
\end{table}

\begin{table*}[th]
\caption{Experimental results on the two datasets. The best results are boldfaced, and the second-best results are underlined. 
}
\label{tab:overall performance}
\vspace{-2mm}
\resizebox{\textwidth}{!}{
\begin{tabular}{c|c|ccccc|cccccc|cc|c|c}
\toprule
Dataset & Metric & BPR   & PinSage & NGCF  & LightGCN & SGL   & NMTR  & MBGCN & MATN  & KHGT  & EHCF  & CML         & ADT         & NoisyTune & DPT            & Imprv. \\ \hline
\multirow{2}{*}{IJCAI} &
  HR &
  0.163 &
  0.176 &
  0.256 &
  0.257 &
  0.249 &
  0.294 &
  0.304 &
  0.369 &
  0.317 &
  0.409 &
  {\ul 0.477} &
  0.475 &
  0.473 &
  \textbf{0.490} &
  2.62\% \\
        & N   & 0.085 & 0.091   & 0.124 & 0.122    & 0.123 & 0.161 & 0.160  & 0.209 & 0.182 & 0.237 & 0.283       & {\ul 0.286} & 0.275     & \textbf{0.294} & 2.65\% \\ \hline
\multirow{2}{*}{Tmall} &
  HR &
  0.243 &
  0.274 &
  0.322 &
  0.342 &
  0.350 &
  0.362 &
  0.381 &
  0.406 &
  0.391 &
  0.433 &
  0.543 &
  0.542 &
  {\ul 0.545} &
  \textbf{0.554} &
  1.58\% \\
        & N   & 0.143 & 0.151   & 0.184 & 0.205    & 0.210  & 0.215 & 0.213 & 0.225 & 0.232 & 0.260  & {\ul 0.327} & 0.324       & 0.325     & \textbf{0.330}  & 0.92\% \\ \bottomrule
\end{tabular}
\vspace{-2mm}
}
\end{table*}

\subsubsection{Baselines.} We compare DPT with various representative recommendation methods, including single- and multi-behavior recommendation models. \textit{Single-behavior recommendation}: (1) \textbf{BPR}~\cite{RFZ09} is a matrix factorization model with the BPR optimization objective.
(2) \textbf{PinSage}~\cite{YHC18} uses an importance-based method to pass the message on paths constructed by a random walk. (3) \textbf{NGCF}~\cite{WHW19} utilizes a standard convolutional message passing method to learn representations. (4) \textbf{LightGCN}~\cite{HDW20} is a lightweight yet effective graph convolution network for representation learning. (5) \textbf{SGL}~\cite{WWF21} performs a self-supervised learning paradigm with multi-view graph augmentation and discrimination. \textit{Multi-behavior recommendation}: (1) \textbf{NMTR}~\cite{GHG19} integrates prior knowledge of behavior relations into a multi-task learning framework. (2) \textbf{MATN}~\cite{XHX20} uses memory-enhanced self-attention mechanism for multi-behavior recommendation. (3) \textbf{MBGCN}~\cite{JGH20} leverages convolutional graph neural network to learn high-order multi-behavioral patterns. (4) \textbf{EHCF}~\cite{CZZ20} conducts knowledge transferring among heterogeneous behaviors and uses a new positive-only loss for model optimization. (5) \textbf{KHGT}~\cite{XHX21} incorporates temporal information and item-side knowledge into the multi-behavior modeling. (6) \textbf{CML}~\cite{WHX22} adopts meta-learning and contrastive learning paradigms to learn distinguishable behavior representations. In addition, we compare DPT with two state-of-the-art lightweight methods by applying them in multi-behavior recommendation: (1) \textbf{ADT}~\cite{WFH21} adaptively prunes noisy interactions for implicit feedback denoising. (2) \textbf{NoisyTune}~\cite{WWQT22} uses a matrix-wise perturbing method to empower the traditional fine-tuning paradigm.

\subsubsection{Implement Details.} Identical to the previous study~\cite{WHX22}, we initialize the trainable parameters with Xavier~\cite{GB10}. The AdamW optimizer~\cite{LH17} and the Cyclical Learning Rate (CLR) strategy~\cite{S17} are adopted with a default base learning rate of $1e^{-3}$ and a max learning rate of $5e^{-3}$. We set the default mini-batch size to $8192$. The dimension $d$ of trainable parameters is set to $16$ and $32$ for IJCAI and Tmall, respectively. The $L_2$ regularization coefficient is searched in $\{1e^{-4},1e^{-3},1e^{-2}\}$. As suggested by CML~\cite{WHX22}, we adopt the dropout operation with a default ratio of $0.8$, and set the maximum layer of the graph encoder to $L=3$ to learn high-order information without falling into over-fitting and over-smoothing issues. Inspired by NoisyTune~\cite{WWQT22}, we set the default disturber $\delta=0.2$ to control noise identification reliability. The hyper-parameters of all competing models either follow the suggestions from the original papers or are carefully tuned, and the best performances are reported. We implement
DPT in PyTorch 1.7.1, Python 3.8.3 on a workstation with an Intel Xeon Platinum 2.40GHz CPU, an NVIDIA Quadro RTX 8000 GPU, and 754GB RAM.

\begin{table}[t]
\caption{Impact of different prompt-tuning methods.}
\vspace{-2mm}
\resizebox{\linewidth}{!}{
\begin{tabular}{l|cc|cc}
\toprule
Dataset     & \multicolumn{2}{c|}{IJCAI}      & \multicolumn{2}{c}{Tmall}       \\ \hline
Metrics     & HR             & NDCG           & HR             & NDCG           \\ \hline
DPT-shallow & 0.472          & 0.274          & 0.536          & 0.316          \\
DPT-projection     & 0.489    &  0.291    &0.548           &0.329           \\
DPT-add     & \textbf{0.490} & \textbf{0.294} & \textbf{0.554} & \textbf{0.330} \\ \bottomrule
\end{tabular}}
\label{tab: different prompt}
\vspace{-2mm}
\end{table}

\subsection{Performance Comparison (RQ1)}
We present the main experimental results in Table~\ref{tab:overall performance}. \textit{Imprv} stands for the average improvements, and all improvements are significant by performing a two-sided $t$-test with $p<0.05$ over the strongest baselines. We can draw a few key observations as follows:
\begin{itemize}[leftmargin=*]
   \item DPT consistently yields the best performance on all datasets. In particular, its relative improvements over the strongest baselines are 
    2.62\% and 1.58\% in terms of HR and 2.65\%, 0.92\% in terms of NDCG on IJCAI and Tmall, respectively. Such results generally demonstrate the superiority of our solution.
    \item Compared with the single-behavior recommendation methods, multi-behavior recommendation models consistently improve performance by a significant margin, which confirms the inherent inadequacy of learning from only a single type of behavior.
    \item Among the multi-behavior methods, DPT consistently achieves the best performance. We attribute such improvements to learning noiseless auxiliary behaviors and bridging the behavioral semantic gap, which can better understand behavior-specific information and thus generate more accurate representations.
    \item Compared with the denoising and fine-tuning methods, DPT can reliably learn inherent noise under auxiliary behaviors and transfer more suitable semantics to target behavior recommendation, demonstrating its effectiveness. 
\end{itemize}

\subsection{Ablation Study (RQ2)}
\label{sec:ablation study}
\begin{figure}[t]
\centering
  \includegraphics[width=\linewidth]{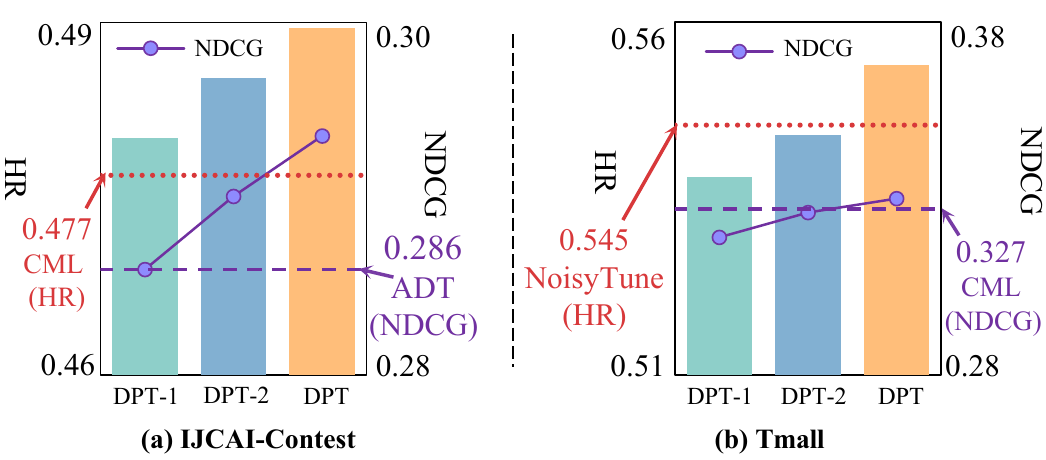}
\vspace{-6mm}
\caption{Impact of different stages of DPT.}
\label{fig:ablation}
\vspace{-3.5mm}
\end{figure}

To verify the contribution of each stage of DPT, we conduct an ablation study with various variants over the two datasets, including (1) \textit{DPT-1} using only the first stage, (2) \textit{DPT-2} using only the first and second stages, and (3) \textit{DPT-3} (or DPT) using all the three stages. Figure~\ref{fig:ablation} shows the performances of different variants in terms of HR and NDCG on IJCAI and Tmall. Red/purple dotted lines represent HR/NDCG of the strongest baselines. It can be observed that each stage positively contributes to performance. With the three-stage learning paradigm, DPT can consistently outperform the other variants. Each stage is better than the previous stage, which validates our motivation that noise and the semantic gap may mislead the target recommendation. \textcolor{black}{It is also worth noting that the forward operations are more efficient than expected. Specifically, DPT-2/3 are 6x/12x faster than DPT-1 per epoch, which confirms the high efficiency of the proposed lightweight tuning approaches. While the first stage is more costly, it is a pre-training process and does not need to be performed frequently.}

We further investigate different variants discussed in Eq.~(\ref{eq:add prompt}) in Table~\ref{tab: different prompt}, including (1) \textit{DPT-shallow} that uses prompt in the first layer of the graph encoder, (2) \textit{DPT-projection} that generates user/item embedding vector projection by the prompt, and (3) \textit{DPT-add} which is our choice. In all cases, DPT-add consistently outperforms the others, confirming the reasonableness of our design choice.

\textcolor{black}{Moreover, we separately remove each type of behaviors to study different behaviors' importance of making recommendations. Specifically, HR (NDCG) @10 on the IJCAI dataset is: w/o click: 0.351 (0.206), w/o add-to-favorite: 0.423 (0.237), w/o add-to-cart: 0.481 (0.285). Such results demonstrate that each type of behaviors contributes to model performance. In particular, clicks are the most important signal. While being noisy, the large number of clicks can contribute the most useful information.}

\subsection{Case Studies of DPT's Explainability (RQ3)}
\label{subsec:case_study}
\begin{figure}[t]
\centering
  \includegraphics[width=\linewidth]{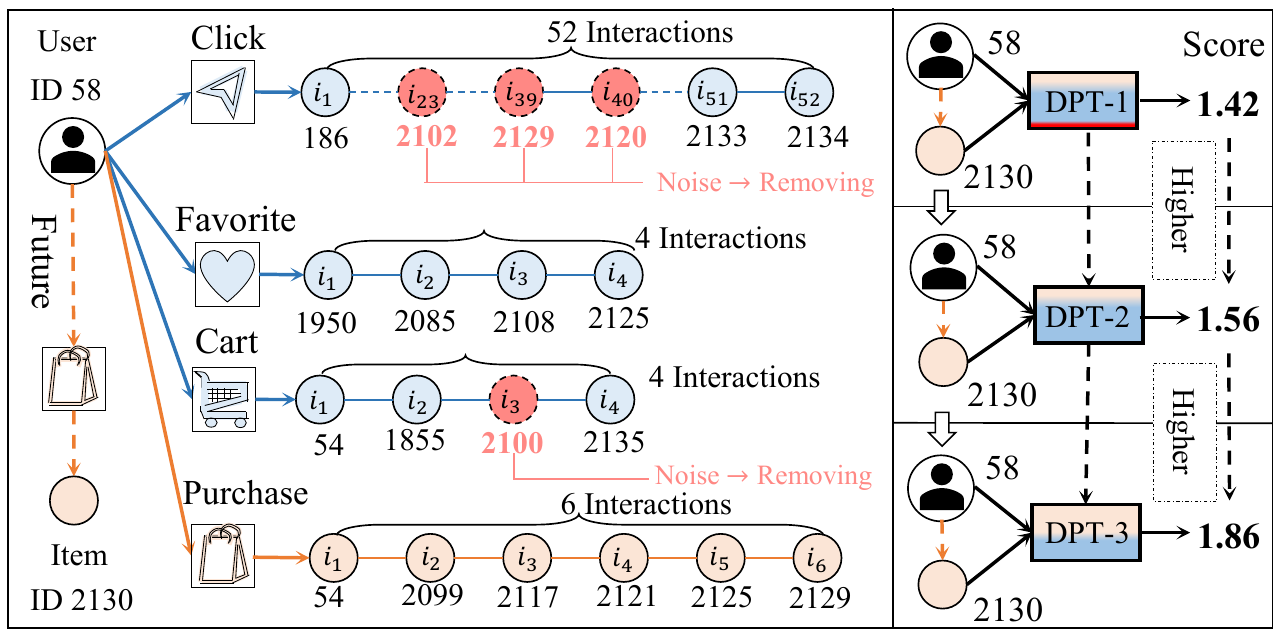}
 \vspace{-5mm}
\caption{A case study to show how the three-stage learning paradigm affects target behavior recommendation.}
\vspace{-2mm}
\label{fig:case-study}
\end{figure}

Finally, we conduct a case study to illustrate how the three-stage learning paradigm of DPT can affect the target behavior recommendation. In Figure~\ref{fig:case-study}, we show a user whose ID is 58 and whose future purchase item ID is 2130. The user has clicked 52 items $\{186, \cdots, 2134\}$, added 4 items $\{1950,\cdots,2125\}$ to favorite, added 4 item $\{54,\cdots, 2135\}$ to cart, and purchased 6 item $\{54,\cdots,2129\}$ from the Tmall dataset. After the first stage of DPT, we explicitly remove noisy item $\{2102,2129,2120,2100\}$ under click and add-to-cart behaviors. After that, we utilize the rest of interactions for the following stages. Each stage of DPT consistently yields higher scores than the prior stage (i.e., scores of 1.42, 1.56, and 1.86 in each stage) to recommend item 2130 to the user under the target behavior. This aligns with our motivation that denoising auxiliary behaviors and bridging the semantic gap can boost the performance of target behavior recommendation. \textcolor{black}{Moreover, after denoising, the interaction ratio and number we drop on Tmall are: click: 1.7\% (18,859), favorite: 0.1\% (38), cart: 0.3\% (352), which shows DPT's capability of eliminating noise.}

\section{Related Work}
\label{sec:related_work}

In view of \textcolor{black}{the limited heterogeneity learning capabilities of} traditional CF methods~\cite{RFZ09,YHC18,WHW19,HDW20,MZX21,WWF21}, which typically assume that users have a single type of behaviors, multi-behavior recommendation studies~\cite{GHG19, GHJ19, XHX20, XHX21, WWQ22,JGH20, CZZ20, XXH21, H21, WHX22} mainly focus on learning distinguishable and representative knowledge from auxiliary behaviors to enhance the relatively sparse target behaviors. We can categorize existing studies into two lines: side-information enhanced (e.g., knowledge-driven~\cite{XHX21,CZZ20}) and behavior-aware balanced methods (e.g., meta-learning~\cite{XXH21,WHX22}). 
The first line proposes to leverage additional information as prior knowledge to learn more informative representations. Despite its effectiveness, the subtle differences among multi-typed behaviors should be tackled carefully, which otherwise causes unbalanced learning of multi-behavior representations. Another line of research resorts to unsupervised learning to improve target behavior recommendation. The intuition is that a standard multi-task learning paradigm is insufficient to learn distinguishing representations under multi-typed behaviors. Therefore, how to adaptively balance the weights of multi-typed behaviors renders a crucial problem. Compared with the first line of research, \textcolor{black}{behavior-aware balanced methods usually perform better since they consider the inherently unbalanced distribution of multi-typed behaviors}. While these two types of methods can outperform the traditional CF approaches due to the consideration of behavior heterogeneity, the numerous uncontrollable auxiliary behaviors may introduce irrelevant (i.e., bridging semantic gap) or noisy (i.e., denoising auxiliary behaviors) information into recommenders, which deserves an in-depth exploration. 

Existing denoising studies tackle the challenge of lacking supervised labels in sequential recommendations by comparing items' relevancy with a target item to explicitly remove irrelevant items ~\cite{QWL21,TWL21,YSS21,SWS21,ZDZ22}, which are unsuitable for multi-behavior recommendation due to a different recommendation purpose. In contrast, some other studies~\cite{WFH21,LZLZ20, ZYZ22} explore how to reduce noise's influence by assigning lower weights to learn representations. However, noisy interactions still exist in auxiliary behaviors and may jeopardize target behavior recommendation performance. 

\textcolor{black}{Prompt-tuning techniques have been widely used in various scenarios (e.g., sequential recommendation~\cite{XPK22}, fair recommendation~\cite{WXY22}, and multiple recommendation tasks~\cite{GLF22}) to incorporate large-scale pre-trained language models (e.g., Transformer) into recommendations. Existing prompt-based recommendation methods typically focus on prompt designing and specific language model tuning, which do not undermine our technical contributions of bridging the semantic gap among multi-typed behaviors. Moreover, the typically required corpus (e.g., user reviews) hinders the adoption of such methods for multi-behavior recommendations.}
In contrast, DPT focuses on pinpointing noise and bridging the semantic gap in unbalanced data without requiring additional labels for multi-behavior recommendation. Thus, DPT can be seamlessly integrated into existing multi-behavior recommendation models.




\section{Conclusion}
\label{sec:conclusion}
In this paper, we studied the problem of multi-behavior recommendation from a new perspective -- how to reduce the negative influences raised by the large amount of auxiliary behaviors on target behavior recommendation. We identified two critical challenges in multi-behavior recommendation: denoising auxiliary behaviors and bridging the semantic gap among multi-typed behaviors. We devised a novel DPT framework with a three-stage learning paradigm to solve the above challenges effectively and efficiently. We conducted comprehensive experiments on multiple datasets to show that our solution can consistently achieve the best performance compared with various state-of-the-art methods.

\begin{acks}
This work was supported by the National Key R\&D Program of China under Grant No. 2020YFB1710200 and the National Natural Science Foundation of China under Grant No. 62072136. Xiangyu Zhao was supported by APRC-CityU New Research Initiatives (No. 9610565, Start-up Grant for New Faculty of City University of Hong Kong), SIRG-CityU Strategic Interdisciplinary Research Grant (No. 7020046, No. 7020074), HKIDS Early Career Research Grant (No. 9360163), Huawei Innovation Research Program and Ant Group (CCF-Ant Research Fund).
\end{acks}

\clearpage
\bibliographystyle{ACM-Reference-Format}
\bibliography{7Reference}


\end{document}